\documentclass[prd,aps,nofootinbib,floatfix,11pt]{revtex4}

\usepackage{amsmath,graphicx,epsfig,amssymb,dsfont,mathtools}
\usepackage[usenames]{color}
\usepackage{ulem} 
\usepackage{bigstrut}
\usepackage{slashed}
\usepackage{multirow}
\usepackage{subfigure}

\allowdisplaybreaks

\begin{document}
	
\title{Semileptonic form factors of $\Xi_{c}\to\Xi$ in QCD sum rules}

\author{
Zhen-Xing Zhao$^{1}$~\footnote{Email: zhaozx19@imu.edu.cn},
Xiao-Yu Sun$^{1}$\footnote{Email: sunxy46@163.com},\\
Fu-Wei Zhang$^{1}$,
Yi-Peng Xing$^{1}$,
Ya-Ting Yang$^{1}$
}

\affiliation{
$^{1}$ School of Physical Science and Technology, \\
Inner Mongolia University, Hohhot 010021, China
}
		
\begin{abstract}
There exists a significant deviation between the most recent Lattice QCD simulation
and experimental measurement by Belle for $\Xi_{c}^{0}\to\Xi^{-}\ell^{+}\nu_{\ell}$.
In this work, we investigate the $\Xi_{c}\to\Xi$ form factors in QCD sum rules.
To this end, the two-point correlation functions of $\Xi_{c}$ and $\Xi$,
and the three-point correlation functions of $\Xi_{c}\to\Xi$ are calculated.
At the QCD level, contributions from up to dimension-6 four-quark operators are considered,
and the leading order results of the Wilson coefficients are obtained.
For the form factors, relatively stable Borel windows can be found.
Our form factors are comparable with those of Lattice QCD, except for $f_{\perp}$.
The obtained form factors are then used to predict the branching ratios
of $\Xi_{c}\to\Xi \ell^{+}\nu_{\ell}$, and our predictions are consistent with the most recent data of ALICE and Belle,
and those of Lattice QCD within error. Given that the branching ratios only contain limited information,
we suggest the experimentalists directly measure the form factors of $\Xi_{c}\to\Xi$.
\end{abstract}

\maketitle

\section{Introduction}

Semileptonic decays can be used to extract CKM matrix elements, which
are important parameters of the standard model (SM). In addition,
lepton flavor universality obtained by studying the semileptonic decays
of different leptonic final states is an important tool to test the
SM. Recently, Belle reported the measurement of the branching ratios
of $\Xi_{c}^{0}\to\Xi^{-}\ell^{+}\nu_{\ell}$ \cite{Li:2021uhk}:
\begin{align}
 & {\cal B}(\Xi_{c}^{0}\to\Xi^{-}e^{+}\nu_{e})=(1.31\pm0.04\pm0.07\pm0.38)\%,\nonumber \\
 & {\cal B}(\Xi_{c}^{0}\to\Xi^{-}\mu^{+}\nu_{\mu})=(1.27\pm0.06\pm0.10\pm0.37)\%,
\end{align}
which are already the highest precision for measuring these processes.
However, most existing theoretical predictions are more or less larger
than these data (see Table \ref{Tab:comparison_semi_lep} below),
including various quark model calculations \cite{Zhao:2018zcb,Faustov:2019ddj,Geng:2020gjh,Ke:2021pxk},
fittings based on SU(3) flavor symmetry \cite{Geng:2017mxn,Geng:2018plk,Geng:2019bfz},
light-cone sum rules (LCSR) analyses \cite{Azizi:2011mw,Aliev:2021wat,Duan:2022yia}.
It is particularly worth pointing out that the most recent Lattice QCD simulation
in Ref. \cite{Zhang:2021oja} shows that:
\begin{align}
 & {\cal B}(\Xi_{c}^{0}\to\Xi^{-}e^{+}\nu_{e})=(2.38\pm0.30\pm0.33)\%,\nonumber \\
 & {\cal B}(\Xi_{c}^{0}\to\Xi^{-}\mu^{+}\nu_{\mu})=(2.29\pm0.29\pm0.31)\%.
\end{align}
One can see that, there is a significant deviation between experimental
measurement and Lattice QCD simulation. Considering the high precision
demonstrated by both, this issue deserves further investigation. 

The authors of Refs. \cite{He:2021qnc,Geng:2022xfz,Geng:2022yxb,Ke:2022gxm}
suggested that, this tension can be resolved by considering the $\Xi_{c}-\Xi_{c}^{\prime}$
mixing on the theoretical side. However, recent Lattice QCD simulation
in Refs. \cite{Liu:2023feb,Liu:2023pwr} and QCD sum rules analysis in
Ref. \cite{Sun:2023noo} have shown that this mixing angle is very small,
only about $1^{\circ}$. Such a small mixing angle clearly cannot resolve the
tension between theory and experiment. The tension still lies there. 

A branching ratio itself contains limited information
after all. We suggest the experimentalists directly measure the
form factors of $\Xi_{c}^{0}\to\Xi^{-}\ell^{+}\nu_{\ell}$, which
can be defined as 
\begin{align}
 & \langle\Xi(p_{2},s_{2})|\bar{s}\gamma_{\mu}(1-\gamma_{5})c|\Xi_{c}(p_{1},s_{1})\rangle\nonumber \\
= & \bar{u}(p_{2},s_{2})\left[\gamma_{\mu}f_{1}(q^{2})+i\sigma_{\mu\nu}\frac{q^{\nu}}{M_{1}}f_{2}(q^{2})+\frac{q_{\mu}}{M_{1}}f_{3}(q^{2})\right]u(p_{1},s_{1})\nonumber \\
- & \bar{u}(p_{2},s_{2})\left[\gamma_{\mu}g_{1}(q^{2})+i\sigma_{\mu\nu}\frac{q^{\nu}}{M_{1}}g_{2}(q^{2})+\frac{q_{\mu}}{M_{1}}g_{3}(q^{2})\right]\gamma_{5}u(p_{1},s_{1}),\label{eq:parametrization_standard}
\end{align}
with $M_{1}=m_{\Xi_{c}}$. In fact, BESIII has performed a similar measurement
for $\Lambda_{c}^{+}\to\Lambda e^{+}\nu_{e}$ in Ref. \cite{BESIII:2022ysa},
where the form factors extracted from experiment is directly compared
with those obtained from Lattice QCD. The comparison between
theory and experiment is sharp and direct, and a very interesting
result was found -- there exists a significant deviation between experimental
measurement and Lattice QCD simulation for the form factors of
$\Lambda_{c}^{+}\to\Lambda e^{+}\nu_{e}$. We can say that Ref.
\cite{BESIII:2022ysa} opened an era of fine comparison.

In this work, we will investigate the form factors of $\Xi_{c}\to\Xi$
in QCD sum rules (QCDSR). At the QCD level, contributions from
up to dimension-6 four-quark operators are considered;
For the Wilson coefficients, the leading order (LO) results are obtained.
QCDSR is a QCD-based approach to deal with
hadronic parameters. It reveals a direct connection between hadron
phenomenology and QCD vacuum structure via a few universal parameters
such as quark condensate and gluon condensate. In Refs. \cite{Shi:2019hbf,Xing:2021enr},
we systematically applied QCD sum rules for the first time to study
the form factors of doubly heavy baryons. To further verify our computing technique,
we also investigated the form factors of $\Lambda_{b}\to\Lambda_{c}$, and found that our
results were comparable with those of experiment, heavy quark effective
theory (HQET) at the next-to-leading power, and Lattice QCD \cite{Zhao:2020mod}. 

The rest of this paper is arranged as follows. In Sec. II, we will
investigate the two-point correlation functions of $\Xi_{c}$ and
$\Xi$ to obtain their pole residues, which are indispensable inputs
when calculating the form factors. At the same time, the continuum threshold parameters,
which are the most important parameters in QCDSR in our opinion, are also determined there.
In Sec. III, we will outline how to extract the form factors of
$\Xi_{c}\to\Xi$ from the three-point correlation functions.
Numerical results of form factors and their phenomenological applications
will be shown in Sec. IV, where our results are also compared with
other theoretical predictions and experimental data. We conclude this
paper in the last section.

\section{Two-point correlation functions and pole residues}

To access the $\Xi_{c}\to\Xi$ form factors, the pole residues and
continuum threshold parameters of initial and final
baryons are indispensable inputs. To this end, in this section we investigate
the two-point correlation functions. In the well-known Ref. \cite{Ioffe:1981kw},
Ioffe, perhaps for the first time, used QCDSR to study the masses of light flavor baryons.
In Ref. \cite{Yang:1993bp}, the authors investigated the neutron-proton mass difference, and
contributions from up to dimension-9 operators were included. For the two-point correlation
functions of heavy flavor baryons, Wang has already done a lot of work,
see, for example, Refs. \cite{Wang:2010fq,Wang:2020mxk}.
We also analyzed the two-point correlation function of
$\Xi_{c}$ in Ref. \cite{Zhao:2020mod}, but did not consider the
contribution of gluon condensate there. For consistency, in this work
we recalculate the two-point correlation functions of $\Xi_{c}$ and $\Xi$,
with the contribution of gluon condensate being considered.

Sum rules start from the definitions of interpolating currents of
hadrons. The following currents are respectively adopted for $\Xi_{c}$
and $\Xi$ \cite{Colangelo:2000dp} 
\begin{align}
J_{\Xi_{Q}} & =\epsilon_{abc}(q_{a}^{T}C\gamma_{5}s_{b})Q_{c},\nonumber \\
J_{\Xi} & =\epsilon_{abc}(s_{a}^{T}C\gamma^{\mu}s_{b})\gamma_{\mu}\gamma_{5}q_{c},\label{eq:interpolating_current}
\end{align}
where $q$ and $Q$ respectively denote a $u/d$ quark and a charm quark, $a,b,c$ are color indices,
and $C$ is the charge conjugation matrix. 

The two-point correlation function is defined as
\begin{equation}
\Pi(p)=i\int d^{4}x\ e^{ip\cdot x}\langle0|T\{J(x)\bar{J}(0)\}|0\rangle.\label{eq:2pt_correlator}
\end{equation}
At the hadron level, after inserting the complete set of hadronic
states, the correlation function in Eq. (\ref{eq:2pt_correlator})
is written into 
\begin{equation}
\Pi^{{\rm had}}(p)=\lambda_{+}^{2}\frac{\slashed p+M_{+}}{M_{+}^{2}-p^{2}}+\lambda_{-}^{2}\frac{\slashed p-M_{-}}{M_{-}^{2}-p^{2}}+\cdots,\label{eq:correlator_had}
\end{equation}
where the contribution from negative-parity baryon is also considered,
and $M_{\pm}$ ($\lambda_{\pm}$) stand for the masses (pole residues)
of positive- and negative-parity baryons.

\begin{figure}[!]
\includegraphics[width=0.6\columnwidth]{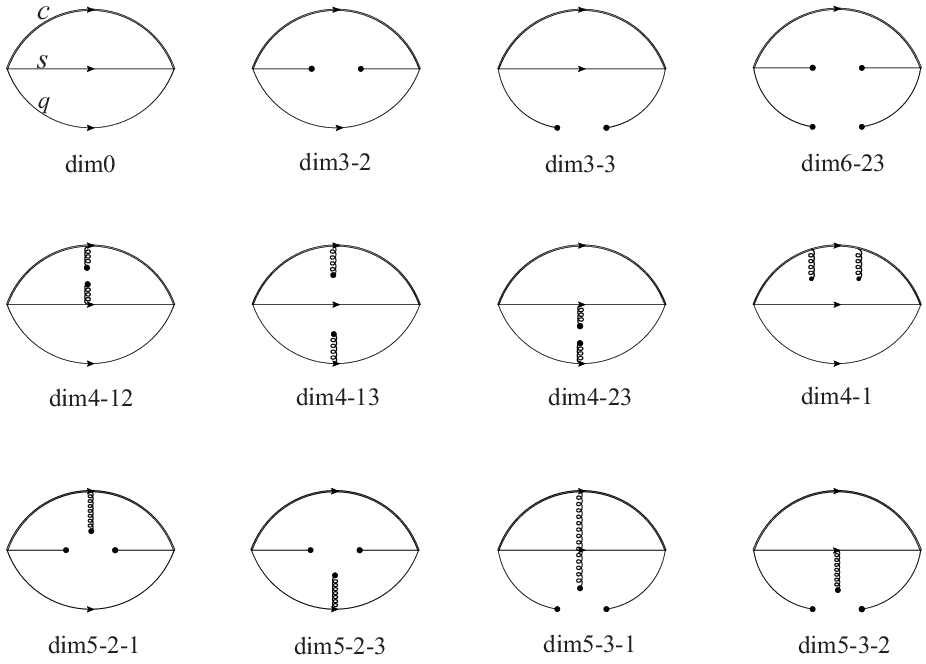}
\caption{All the diagrams considered for the two-point correlation function of $\Xi_{c}$
at the QCD level.}
\label{fig:diagrams_2pt_Xic} 
\end{figure}

\begin{figure}[!]
\includegraphics[width=0.8\columnwidth]{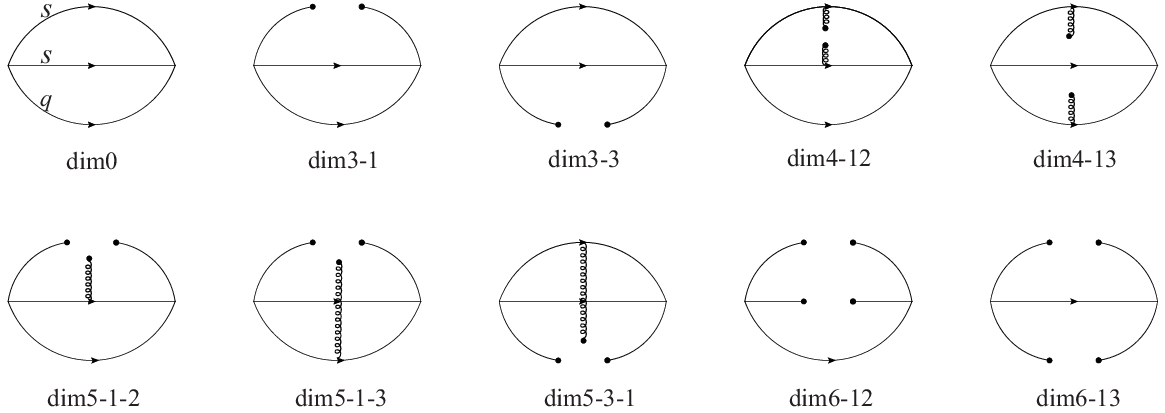}
\caption{All ``independent'' diagrams considered for the two-point correlation function of $\Xi$
at the QCD level. Here ``independent'' means that equivalent diagrams are not shown here.}
\label{fig:diagrams_2pt_Xi} 
\end{figure}

At the QCD level, the correlation function in Eq. (\ref{eq:2pt_correlator})
can be calculated using OPE technique. In this work, contributions
from up to dimension-6 four-quark operators are considered and the
corresponding diagrams for $\Xi_{c}$ and $\Xi$ can be found in Fig. \ref{fig:diagrams_2pt_Xic}
and Fig. \ref{fig:diagrams_2pt_Xi}, respectively.
The calculation results of the correlation function at the QCD level can be formally written as 
\begin{equation}
\Pi^{{\rm QCD}}(p)=A(p^{2})\slashed p+B(p^{2}),\label{eq:2pt_correlator_QCD}
\end{equation}
where the coefficient functions $A$ and $B$ can be further written
into the following dispersion integrals for practical purpose
\begin{equation}
A(p^{2})=\int ds\ \frac{\rho^{A}(s)}{s-p^{2}},\quad B(p^{2})=\int ds\ \frac{\rho^{B}(s)}{s-p^{2}}.\label{eq:rho_A_rho_B}
\end{equation}


Taking advantage of quark-hadron duality and then performing the Borel
transform, one can arrive at the sum rule for the $1/2^{+}$ baryon 
\begin{align}
(M_{+}+M_{-})\lambda_{+}^{2}\exp(-M_{+}^{2}/T_{+}^{2}) & =\int^{s_{+}}ds\ (M_{-}\rho^{A}+\rho^{B})\exp(-s/T_{+}^{2}),\label{eq:2pt_sum_rule}
\end{align}
where $T_{+}^{2}$ and $s_{+}$ are the Borel parameter and continuum
threshold parameter, respectively. Differentiating Eq. (\ref{eq:2pt_sum_rule})
with respect to $-1/T_{+}^{2}$, one can obtain the mass of the $1/2^{+}$
baryon 
\begin{equation}
M_{+}^{2}=\frac{\int^{s_{+}}ds\ (M_{-}\rho^{A}+\rho^{B})\ s\ \exp(-s/T_{+}^{2})}{\int^{s_{+}}ds\ (M_{-}\rho^{A}+\rho^{B})\ \exp(-s/T_{+}^{2})}.\label{eq:mass_sum_rule}
\end{equation}
In this work, Eq. (\ref{eq:mass_sum_rule}) is viewed as a constraint
of Eq. (\ref{eq:2pt_sum_rule}), and is used to fix the continuum
threshold parameter $s_{+}$, which is the most important parameter in QCDSR in our opinion. 

\section{Three-point correlation functions and form factors}

\label{sec:3pt}

In practice, the following simpler parametrization is adopted to extract
the analytical expressions of $\Xi_{c}\to\Xi$ transition form factors:
\begin{align}
 & \langle{\cal B}_{2}(p_{2},s_{2})|\bar{s}\gamma_{\mu}(1-\gamma_{5})c|{\cal B}_{1}(p_{1},s_{1})\rangle\nonumber \\
= & \bar{u}(p_{2},s_{2})\left[\frac{p_{1\mu}}{M_{1}}F_{1}(q^{2})+\frac{p_{2\mu}}{M_{2}}F_{2}(q^{2})+\gamma_{\mu}F_{3}(q^{2})\right]u(p_{1},s_{1})\nonumber \\
- & \bar{u}(p_{2},s_{2})\left[\frac{p_{1\mu}}{M_{1}}G_{1}(q^{2})+\frac{p_{2\mu}}{M_{2}}G_{2}(q^{2})+\gamma_{\mu}G_{3}(q^{2})\right]\gamma_{5}u(p_{1},s_{1}),\label{eq:parametrization_simple}
\end{align}
where ${\cal B}_{1,2}$ denote $\Xi{}_{c}$ and $\Xi$, respectively.
The form factors $F_{i}$ and $G_{i}$ are related to $f_{i}$ and
$g_{i}$ defined in Eq. (\ref{eq:parametrization_standard}) through 
\begin{align}
 & F_{1}=f_{2}+f_{3},\quad F_{2}=\frac{M_{2}}{M_{1}}(f_{2}-f_{3}),\quad F_{3}=f_{1}-\frac{M_{1}+M_{2}}{M_{1}}f_{2};\nonumber \\
 & G_{1}=g_{2}+g_{3},\quad G_{2}=\frac{M_{2}}{M_{1}}(g_{2}-g_{3}),\quad G_{3}=g_{1}+\frac{M_{1}-M_{2}}{M_{1}}g_{2}.
\end{align}
In addition, helicity form factors are usually adopted
by Lattice QCD \cite{Zhang:2021oja,Detmold:2015aaa}, and are related to the form
factors in Eq. (\ref{eq:parametrization_standard}) as follows
\begin{align}
 & f_{+}=f_{1}-\frac{q^{2}}{M_{1}(M_{1}+M_{2})}f_{2},\quad f_{\perp}=f_{1}-\frac{M_{1}+M_{2}}{M_{1}}f_{2},\quad f_{0}=f_{1}+\frac{q^{2}}{M_{1}(M_{1}-M_{2})}f_{3},\nonumber \\
 & g_{+}=g_{1}+\frac{q^{2}}{M_{1}(M_{1}-M_{2})}g_{2},\quad g_{\perp}=g_{1}+\frac{M_{1}-M_{2}}{M_{1}}g_{2},\quad g_{0}=g_{1}-\frac{q^{2}}{M_{1}(M_{1}+M_{2})}g_{3}.
\end{align}
In this work, the results of these helicity form factors are presented to make a close comparison
with those of Lattice QCD.

The following three-point correlation functions are defined to extract
the form factors of $\Xi_{c}\to\Xi$ 
\begin{equation}
\Pi_{\mu}^{V,A}(p_{1},p_{2})=i^{2}\int d^{4}xd^{4}y\ e^{-ip_{1}\cdot x+ip_{2}\cdot y}\langle0|T\{J_{\Xi}(y)(V_{\mu},A_{\mu})(0)\bar{J}_{\Xi_{c}}(x)\}|0\rangle,\label{eq:3pt_correlators}
\end{equation}
where $V_{\mu}(A_{\mu})=\bar{s}\gamma_{\mu}(\gamma_{\mu}\gamma_{5})c$
is the vector (axial-vector) current for the $c\to s$ process. The
correlation functions are then calculated at the hadron level and
QCD level. 

At the hadron level, after inserting the complete sets of initial and
final states and considering the contributions from negative-parity
baryons, the vector current correlation function in Eq. (\ref{eq:3pt_correlators})
can be written into 
\begin{eqnarray}
\Pi_{\mu}^{V,{\rm had}}(p_{1},p_{2}) & = & \lambda_{f}^{+}\lambda_{i}^{+}\frac{(\slashed p_{2}+M_{2}^{+})(\frac{p_{1\mu}}{M_{1}^{+}}F_{1}^{++}+\frac{p_{2\mu}}{M_{2}^{+}}F_{2}^{++}+\gamma_{\mu}F_{3}^{++})(\slashed p_{1}+M_{1}^{+})}{(p_{2}^{2}-M_{2}^{+2})(p_{1}^{2}-M_{1}^{+2})}\nonumber \\
 & + & \lambda_{f}^{+}\lambda_{i}^{-}\frac{(\slashed p_{2}+M_{2}^{+})(\frac{p_{1\mu}}{M_{1}^{-}}F_{1}^{+-}+\frac{p_{2\mu}}{M_{2}^{+}}F_{2}^{+-}+\gamma_{\mu}F_{3}^{+-})(\slashed p_{1}-M_{1}^{-})}{(p_{2}^{2}-M_{2}^{+2})(p_{1}^{2}-M_{1}^{-2})}\nonumber \\
 & + & \lambda_{f}^{-}\lambda_{i}^{+}\frac{(\slashed p_{2}-M_{2}^{-})(\frac{p_{1\mu}}{M_{1}^{+}}F_{1}^{-+}+\frac{p_{2\mu}}{M_{2}^{-}}F_{2}^{-+}+\gamma_{\mu}F_{3}^{-+})(\slashed p_{1}+M_{1}^{+})}{(p_{2}^{2}-M_{2}^{-2})(p_{1}^{2}-M_{1}^{+2})}\nonumber \\
 & + & \lambda_{f}^{-}\lambda_{i}^{-}\frac{(\slashed p_{2}-M_{2}^{-})(\frac{p_{1\mu}}{M_{1}^{-}}F_{1}^{--}+\frac{p_{2\mu}}{M_{2}^{-}}F_{2}^{--}+\gamma_{\mu}F_{3}^{--})(\slashed p_{1}-M_{1}^{-})}{(p_{2}^{2}-M_{2}^{-2})(p_{1}^{2}-M_{1}^{-2})}\nonumber \\
 & + & \cdots.\label{eq:3pt_correlator_had}
\end{eqnarray}
In Eq. (\ref{eq:3pt_correlator_had}), $M_{1(2)}^{+(-)}$ denotes
the mass of initial (final) positive- (negative-) parity baryon, and
$F_{1}^{+-}$ is the form factor $F_{1}$ with the positive-parity
final state and negative-parity initial state, and so forth. To arrive
at Eq. (\ref{eq:3pt_correlator_had}), we have adopted the following
definitions of pole residues for positive- and negative-parity baryons
\begin{align}
\langle0|J_{+}|{\cal B}_{+}(p,s)\rangle & =\lambda_{+}u(p,s),\nonumber \\
\langle0|J_{+}|{\cal B}_{-}(p,s)\rangle & =(i\gamma_{5})\lambda_{-}u(p,s),\label{eq:pole_residue}
\end{align}
and the following conventions for the 12 form factors $F_{i}^{\pm\pm}$
\begin{align}
\langle{\cal B}_{f}^{+}(p_{2},s_{2})|V_{\mu}|{\cal B}_{i}^{+}(p_{1},s_{1})\rangle & =\bar{u}_{{\cal B}_{f}^{+}}(p_{2},s_{2})[\frac{p_{1\mu}}{M_{1}^{+}}F_{1}^{++}+\frac{p_{2\mu}}{M_{2}^{+}}F_{2}^{++}+\gamma_{\mu}F_{3}^{++}]u_{{\cal B}_{i}^{+}}(p_{1},s_{1}),\nonumber \\
\langle{\cal B}_{f}^{+}(p_{2},s_{2})|V_{\mu}|{\cal B}_{i}^{-}(p_{1},s_{1})\rangle & =\bar{u}_{{\cal B}_{f}^{+}}(p_{2},s_{2})[\frac{p_{1\mu}}{M_{1}^{-}}F_{1}^{+-}+\frac{p_{2\mu}}{M_{2}^{+}}F_{2}^{+-}+\gamma_{\mu}F_{3}^{+-}](i\gamma_{5})u_{{\cal B}_{i}^{-}}(p_{1},s_{1}),\nonumber \\
\langle{\cal B}_{f}^{-}(p_{2},s_{2})|V_{\mu}|{\cal B}_{i}^{+}(p_{1},s_{1})\rangle & =\bar{u}_{{\cal B}_{f}^{-}}(p_{2},s_{2})(i\gamma_{5})[\frac{p_{1\mu}}{M_{1}^{+}}F_{1}^{-+}+\frac{p_{2\mu}}{M_{2}^{-}}F_{2}^{-+}+\gamma_{\mu}F_{3}^{-+}]u_{{\cal B}_{i}^{+}}(p_{1},s_{1}),\nonumber \\
\langle{\cal B}_{f}^{-}(p_{2},s_{2})|V_{\mu}|{\cal B}_{i}^{-}(p_{1},s_{1})\rangle & =\bar{u}_{{\cal B}_{f}^{-}}(p_{2},s_{2})(i\gamma_{5})[\frac{p_{1\mu}}{M_{1}^{-}}F_{1}^{--}+\frac{p_{2\mu}}{M_{2}^{-}}F_{2}^{--}+\gamma_{\mu}F_{3}^{--}](i\gamma_{5})u_{{\cal B}_{i}^{-}}(p_{1},s_{1}).
\end{align}

\begin{figure}[!]
\includegraphics[width=0.9\columnwidth]{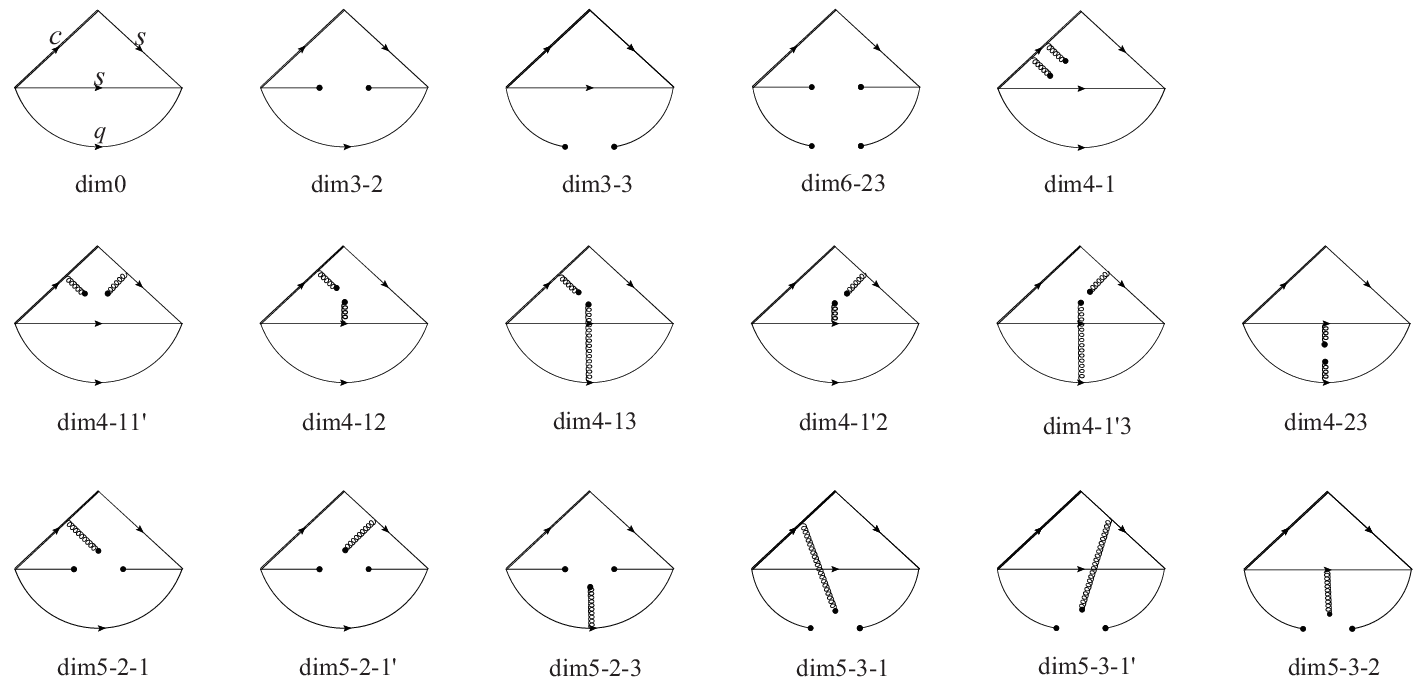}
\caption{All the diagrams considered for the three-point correlation functions of $\Xi_{c}\to\Xi$
at the QCD level.}
\label{fig:diagrams_3pt} 
\end{figure}

At the QCD level, contributions from up to dimension-6 four-quark operators
are considered, as can be seen in Fig. \ref{fig:diagrams_3pt}. The calculation results of the
vector current correlation function in Eq. (\ref{eq:3pt_correlators})
can be formally written as 
\begin{equation}
\Pi_{\mu}^{V,{\rm QCD}}(p_{1},p_{2})=\sum_{i=1}^{12}A_{i}e_{i\mu}\label{eq:3pt_correlator_QCD_formal}
\end{equation}
with 
\begin{eqnarray}
(e_{1,2,3,4})_{\mu} & = & \{\slashed p_{2},1\}\times\{p_{1\mu}\}\times\{\slashed p_{1},1\},\nonumber \\
(e_{5,6,7,8})_{\mu} & = & \{\slashed p_{2},1\}\times\{p_{2\mu}\}\times\{\slashed p_{1},1\},\nonumber \\
(e_{9,10,11,12})_{\mu} & = & \{\slashed p_{2},1\}\times\{\gamma_{\mu}\}\times\{\slashed p_{1},1\}.\label{eq:e_i_mu}
\end{eqnarray}
The coefficients $A_{i}$ in Eq. (\ref{eq:3pt_correlator_QCD_formal})
are then expressed as double dispersion integrals 
\begin{equation}
A_{i}(p_{1}^{2},p_{2}^{2},q^{2})=\int^{\infty}ds_{1}\int^{\infty}ds_{2}\frac{\rho_{i}(s_{1},s_{2},q^{2})}{(s_{1}-p_{1}^{2})(s_{2}-p_{2}^{2})},
\end{equation}
where the spectral functions $\rho_{i}(s_{1},s_{2},q^{2})$ can be obtained
by applying Cutkosky cutting rules to the diagrams in Fig. \ref{fig:diagrams_3pt}.

Equating Eqs. (\ref{eq:3pt_correlator_had}) and (\ref{eq:3pt_correlator_QCD_formal}),
and using the quark-hadron duality, one can arrive at 12 equations
for 12 unknown form factors $F_{i}^{\pm\pm}$. Solving these equations,
and then performing the Borel transform, one can obtain the following
expressions for $F_{i}^{++}$: 
\begin{align}
\lambda_{i}^{+}\lambda_{f}^{+}(F_{1}^{++}/M_{1}^{+})\exp\left(-\frac{M_{1}^{+2}}{T_{1}^{2}}-\frac{M_{2}^{+2}}{T_{2}^{2}}\right) & =\frac{\{M_{1}^{-}M_{2}^{-},M_{2}^{-},M_{1}^{-},1\}.\{{\cal B}A_{1},{\cal B}A_{2},{\cal B}A_{3},{\cal B}A_{4}\}}{(M_{1}^{+}+M_{1}^{-})(M_{2}^{+}+M_{2}^{-})},\nonumber \\
\lambda_{i}^{+}\lambda_{f}^{+}(F_{2}^{++}/M_{2}^{+})\exp\left(-\frac{M_{1}^{+2}}{T_{1}^{2}}-\frac{M_{2}^{+2}}{T_{2}^{2}}\right) & =\frac{\{M_{1}^{-}M_{2}^{-},M_{2}^{-},M_{1}^{-},1\}.\{{\cal B}A_{5},{\cal B}A_{6},{\cal B}A_{7},{\cal B}A_{8}\}}{(M_{1}^{+}+M_{1}^{-})(M_{2}^{+}+M_{2}^{-})},\nonumber \\
\lambda_{i}^{+}\lambda_{f}^{+}F_{3}^{++}\exp\left(-\frac{M_{1}^{+2}}{T_{1}^{2}}-\frac{M_{2}^{+2}}{T_{2}^{2}}\right) & =\frac{\{M_{1}^{-}M_{2}^{-},M_{2}^{-},M_{1}^{-},1\}.\{{\cal B}A_{9},{\cal B}A_{10},{\cal B}A_{11},{\cal B}A_{12}\}}{(M_{1}^{+}+M_{1}^{-})(M_{2}^{+}+M_{2}^{-})},\label{eq:Fi_plus_plus}
\end{align}
where 
\begin{equation}
{\cal B}A_{i}\equiv\int^{s_{10}}ds_{1}\int^{s_{20}}ds_{2}\ \rho_{i}(s_{1},s_{2},q^{2})\exp(-s_{1}/T_{1}^{2})\exp(-s_{2}/T_{2}^{2}),
\end{equation}
are doubly Borel transformed coefficients, with $s_{1(2)}^{0}$ the
continuum threshold parameter of the initial (final) baryon, and $T_{1,2}^{2}$ are the Borel parameters.

\subsection{The leading logarithmic corrections}

In this work, we also consider the leading logarithmic (LL) corrections
for the pole residues and form factors. For this purpose, in the following,
we will first briefly summarize some key points of the operator product expansion (OPE) technique.

For two operators ${\cal O}_{1}$ and ${\cal O}_{2}$ separated by
a small distance $x$, the product of these two operators can be computed
using OPE
\begin{equation}
{\cal O}_{1}(x){\cal O}_{2}(0)=\sum_{n}C_{n}(x){\cal O}_{n}(0),\label{eq:OPE_2O}
\end{equation}
where ${\cal O}_{1,2,n}$ are defined at some renormalizaion scale
$\mu$. The calculated Wilson coefficient $C_{n}$ should be multiplied
by a LL correction factor \cite{Peskin:1995ev}
\begin{equation}
\left(\frac{\log(1/|x|^{2}\Lambda_{{\rm QCD}}^{2})}{\log(\mu^{2}/\Lambda_{{\rm QCD}}^{2})}\right)^{(a_{n}-a_{1}-a_{2})/2\beta_{0}},\label{eq:LL_factor}
\end{equation}
where $a_{{\cal O}}$ is related to the anomalous dimension $\gamma_{{\cal O}}$
by
\begin{equation}
\gamma_{{\cal O}}=-a_{{\cal O}}\frac{g^{2}}{(4\pi)^{2}},
\end{equation}
and $\beta_{0}$ is the first coefficient of the QCD $\beta$ function
\begin{equation}
\beta_{0}=11-\frac{2}{3}n_{f}.
\end{equation}
Note that, after performing the Fourier transform as in Eqs. (\ref{eq:2pt_correlator})
and (\ref{eq:3pt_correlators}), the inverse of the squared distance
$1/|x|^{2}$ is actually $\sim p^{2}$. 

In Ref. \cite{Zhao:2021xwl}, we explicitly calculated the LO
anomalous dimensions of the interpolating currents in Eq. (\ref{eq:interpolating_current}),
and found that the two anomalous dimensions happen to be the same,
both equal to
\begin{equation}
\gamma_{J}=-4\frac{g^{2}}{(4\pi)^{2}}.
\end{equation}
The anomalous dimension of $\bar{\psi}\psi$ can be found in any standard
quantum field theory textbook
\begin{equation}
\gamma_{\bar{\psi}\psi}=-8\frac{g^{2}}{(4\pi)^{2}}.
\end{equation}
Following Ioffe in Ref. \cite{Ioffe:1981kw}, the LL corrections of the
Wilson coefficients for higher-dimensional operators are no longer
considered due to the following reasons:
\begin{itemize}
\item The contribution of these terms is comparatively small.
\item The numerical values of higher-dimensional condensate parameters contain
large ambiguity. 
\end{itemize}
Some remarks on the OPE of three-point operator product
\begin{equation}
{\cal O}_{1}(y){\cal O}_{2}(0){\cal O}_{3}(x)=\sum_{n}C_{n}(x,y){\cal O}_{n}(0)\label{eq:OPE_3O}
\end{equation}
are in order. If Eq. (\ref{eq:OPE_3O}) is considered to have been
expanded twice using Eq. (\ref{eq:OPE_2O}), one can easily check
that in the limit of
\begin{equation}
|x|=|y|,\label{eq:equal_distance}
\end{equation}
the corresponding LL correction factor, similar to that in Eq. (\ref{eq:LL_factor}), is
\begin{equation}
\left(\frac{\log(1/|x|^{2}\Lambda_{{\rm QCD}}^{2})}{\log(\mu^{2}/\Lambda_{{\rm QCD}}^{2})}\right)^{(a_{n}-a_{1}-a_{2}-a_{3})/2\beta_{0}}.\label{eq:LL_factor_2}
\end{equation}
For the $B\to\pi$ process, the approximation in Eq. (\ref{eq:LL_factor_2})
is bad; However, as long as the mass difference between the initial
and final states is not very large, this approximation should not be bad.
$\Xi_{c}\to\Xi$ can be attributed to the latter situation. 

\section{Numerical results and phenomenological applications}

Numerical results will be shown in this section, and our main results include the pole residues,
the continuum threshold parameters of $\Xi_{c}$ and $\Xi$, and the $\Xi_{c}\to\Xi$ form factors.
The main inputs include the condensate parameters and quark masses. 
The condensate parameters are taken as \cite{Colangelo:2000dp}:
$\langle\bar{q}q\rangle(1\ {\rm GeV})=-(0.24\pm0.01\ {\rm GeV})^{3}$,
$\langle\bar{s}s\rangle=(0.8\pm0.2)\langle\bar{q}q\rangle$, $\langle g_{s}^{2}G^{2}\rangle=(0.47\pm0.14)\ {\rm GeV}^{4}$,
$\langle\bar{q}g_{s}\sigma Gq\rangle=m_{0}^{2}\langle\bar{q}q\rangle$
and $\langle\bar{s}g_{s}\sigma Gs\rangle=m_{0}^{2}\langle\bar{s}s\rangle$
with $m_{0}^{2}=(0.8\pm0.2)\ {\rm GeV}^{2}$. The following quark
masses are adopted \cite{ParticleDataGroup:2022pth}: 
\begin{align}
 & m_{c}(m_{c})=1.27\pm0.02\ {\rm GeV},\quad m_{s}(2\ {\rm GeV})=0.093\pm0.009\ {\rm GeV},
\end{align}
and $m_{u/d}$ is taken to be 0. 

When calculating the pole residues of $\Xi_{c}$ and $\Xi$, and the
form factors of $\Xi_{c}\to\Xi$, we take all the renormalization
scales at $\mu=m_{c}$. The following equation for the QCD running
coupling constant at the one-loop level has been used
\begin{equation}
\alpha_{s}(\mu)=\frac{4\pi}{\beta_{0}\log(\mu^{2}/\Lambda_{{\rm QCD}}^{2})},
\end{equation}
and $\alpha_{s}(m_{Z})=0.118$ \cite{ParticleDataGroup:2022pth} is taken as a reference point for renormalization.
The continuity of $\alpha_{s}$ allow to find values of $\Lambda_{{\rm QCD}}^{(n_{f})}$
for different $n_{f}$. It turns out that: $\Lambda_{{\rm QCD}}^{(5)}=88\ {\rm MeV}$,
$\Lambda_{{\rm QCD}}^{(4)}=120\ {\rm MeV}$, and $\Lambda_{{\rm QCD}}^{(3)}=143\ {\rm MeV}$.
Especially, $\alpha_{s}(m_{c})\approx0.32$ can be obtained. 
Then, the quark masses and condensate parameters can be evolved through
their respective one-loop evolution formulas. For example, for the
quark mass, the one-loop evolution formula is 
\begin{equation}
m(\mu_{2})=m(\mu_{1})\left(\frac{\log(\mu_{1}^{2}/\Lambda_{{\rm QCD}}^{2})}{\log(\mu_{2}^{2}/\Lambda_{{\rm QCD}}^{2})}\right)^{4/\beta_{0}}.
\end{equation}

As can be seen in Figs. \ref{fig:diagrams_2pt_Xic}, \ref{fig:diagrams_2pt_Xi},
and \ref{fig:diagrams_3pt} that, we have only considered the tree-level
diagrams -- when cutting rules are applied, there is no longer a
loop diagram. For the Wilson coefficients, we have only obtained the
LO results. As can be seen in our previous work \cite{Zhao:2021lzd},
the error caused by scale dependence plays an important
role. To reduce the dependence of calculation results on the renormalization
scale, in this work, we also consider the LL corrections
to the Wilson coefficients. However, numerically these corrections
are small, the reason is given as follows.

In Eq. (\ref{eq:LL_factor}) and Eq. (\ref{eq:LL_factor_2}), the
renormalization scale $\mu=m_{c}$, and the inverse of the distance
$1/|x|$ is exactly or close to ${\cal O}(m_{c})$ for the two-point
correlation functions of $\Xi_{c}$ and $\Xi$, and the three-point
correlation functions of $\Xi_{c}\to\Xi$. Therefore, all the LL correction
factors are all close to $1$, that is, the LL
corrections are small (only a few percent). 

How to evaluate the contribution from the next-to-leading order (NLO)? This
is certainly a difficult question to answer, and often only through
detailed calculations at NLO can a clear answer be obtained. The calculation
of the decay constants $f_{B_{(s)}}$ in Ref. \cite{Jamin:2001fw} are excepted
to shed some light on the contributions from higher orders.
As pointed out in Ref. \cite{Jamin:2001fw}, in the pole mass scheme, the convergence
is poor, while in the $\overline{{\rm MS}}$ scheme, the convergence
is good. In the $\overline{{\rm MS}}$ scheme, the contribution from
NLO is about 10\%. Of course, this is only for the bottom quark case
and also limited to the two-point correlation functions. The NLO correction
for the charm quark case should be larger. 

Inspired by the pole mass scheme, in this work, we also take $m_{c}$
and $m_{s}$ as the pole masses, and expand them to the NLO \cite{ParticleDataGroup:2022pth}
\begin{equation}
m^{{\rm pole}}=m(\mu)\left(1+\frac{4\alpha_{s}(\mu)}{3\pi}\right)
\end{equation}
to evaluate the contribution from NLO. That is, in this work, two sets of results will be presented
\begin{itemize}
\item LO + LL + $\overline{{\rm MS}}$ mass, which is taken as the central value;
\item LO + LL + pole mass@NLO, which is used to evaluate the contribution from
NLO.
\end{itemize}
We find that, there are respectively about 20\%, 15\% uncertainties
between these two schemes, for the pole residues of $\Xi_{c}$, and the form factors of $\Xi_{c}\to\Xi$.
These numbers more or less meet the expectation above.
However, it is worth emphasizing again that this is only a very rough
estimate, and more accurate numbers can only be known after
performing the calculation of  NLO.

\subsection{The pole residues of $\Xi_{c}$ and $\Xi$}

\begin{figure}[!]
\includegraphics[width=1.0\columnwidth]{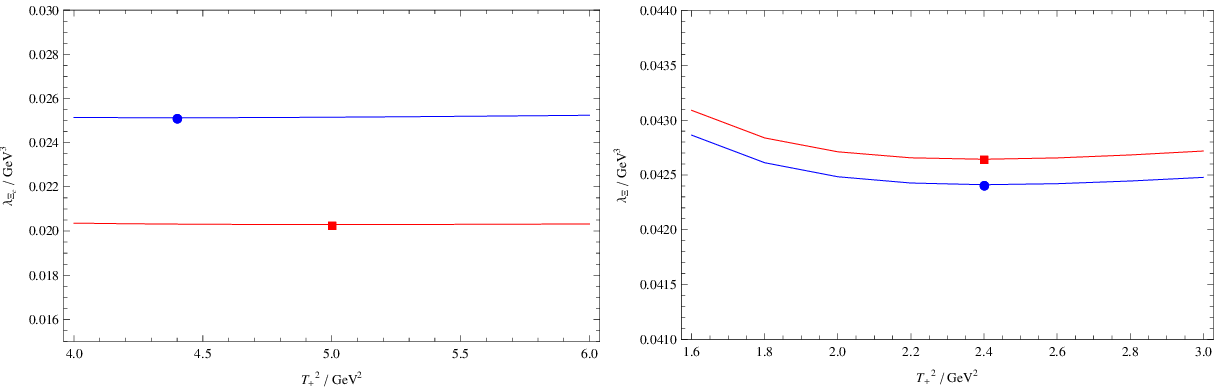}
\caption{The pole residues of $\Xi_{c}$ and $\Xi$ as functions of the Borel parameter $T_{+}^{2}$.
The blue and red curves correspond to the $\overline{{\rm MS}}$ scheme and the pole mass scheme, respectively.
The extreme points on these curves correspond to the experimental value of the baryon mass.
The $s_{+}$ and $T_{+}^{2}$ corresponding to these extreme points can be found in Table \ref{Tab:pole_residues}.}
\label{fig:pole_residues} 
\end{figure}

\begin{table}
\caption{Our predictions of the pole residues of $\Xi_{c}$ and $\Xi$. Optimal
$s_{+}$ and $T_{+}^{2}$ are also shown.}
\label{Tab:pole_residues} %
\begin{tabular}{c|c|c||c|c|cc}
\hline 
$\Xi_{c}$  & $(s_{+}/{\rm GeV}^{2},\ T_{+}^{2}/{\rm GeV}^{2})$  & $\lambda_{+}/{\rm GeV}^{3}$  & $\Xi$  & $(s_{+}/{\rm GeV}^{2},\ T_{+}^{2}/{\rm GeV}^{2})$  & $\lambda_{+}/{\rm GeV}^{3}$  & \tabularnewline
\hline 
$\overline{{\rm MS}}$ mass  & $(8.74,4.4)$  & $0.0251$  & $\overline{{\rm MS}}$ mass  & $(3.22,2.4)$  & $0.0424$  & \tabularnewline
\hline 
Pole mass@NLO  & $(8.70,5.0)$  & $0.0203$  & Pole mass@NLO  & $(3.24,2.4)$  & $0.0426$  & \tabularnewline
\hline 
\end{tabular}
\end{table}

The pole residues of $\Xi_{c}$ and $\Xi$ are determined using the
sum rule in Eq. (\ref{eq:2pt_sum_rule}) using Eq. (\ref{eq:mass_sum_rule}) as a constraint,
and the corresponding results
are shown in Fig. \ref{fig:pole_residues} and Table \ref{Tab:pole_residues}. Here is one comment.
The experimental masses of $\Xi_{c}^{+,0}$
are respectively $2.468$ GeV and $2.470$ GeV, while those
of $\Xi^{0,-}$ are respectively $1.315$ GeV
and $1.322$ GeV \cite{ParticleDataGroup:2022pth}. In Fig. \ref{fig:pole_residues}
and Table \ref{Tab:pole_residues}, we have actually used the experimental
masses of $\Xi_{c}^{0}$ and $\Xi^{-}$. Note that our QCDSR analysis
is blind to $u$ or $d$ quark within the hadron. 

\subsection{The form factors of $\Xi_{c}\to\Xi$}

\begin{figure}[!]
\includegraphics[width=1.0\columnwidth]{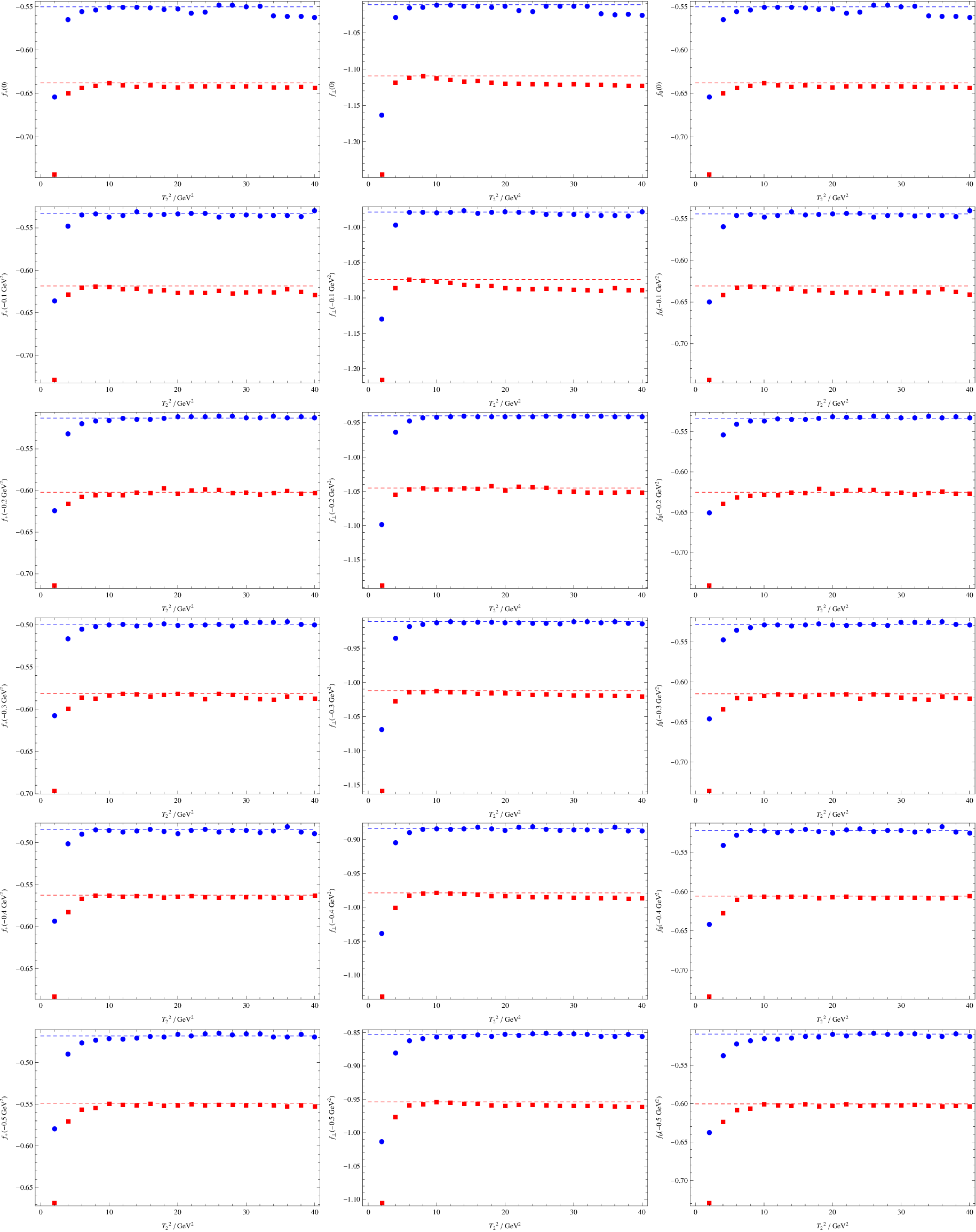}
\caption{The helicity form factors $f_{+,\perp,0}(q^{2})$ as functions of the Borel parameter $T_{2}^{2}$.
The blue dots and red squares respectively correspond to the results obtained using the $\overline{{\rm MS}}$ scheme
and the pole mass scheme. $q^{2}=0.0, -0.1, ..., -0.5\ {\rm GeV}^{2}$ are considered. }
\label{fig:FF_V} 
\end{figure}
\begin{figure}[!]
\includegraphics[width=1.0\columnwidth]{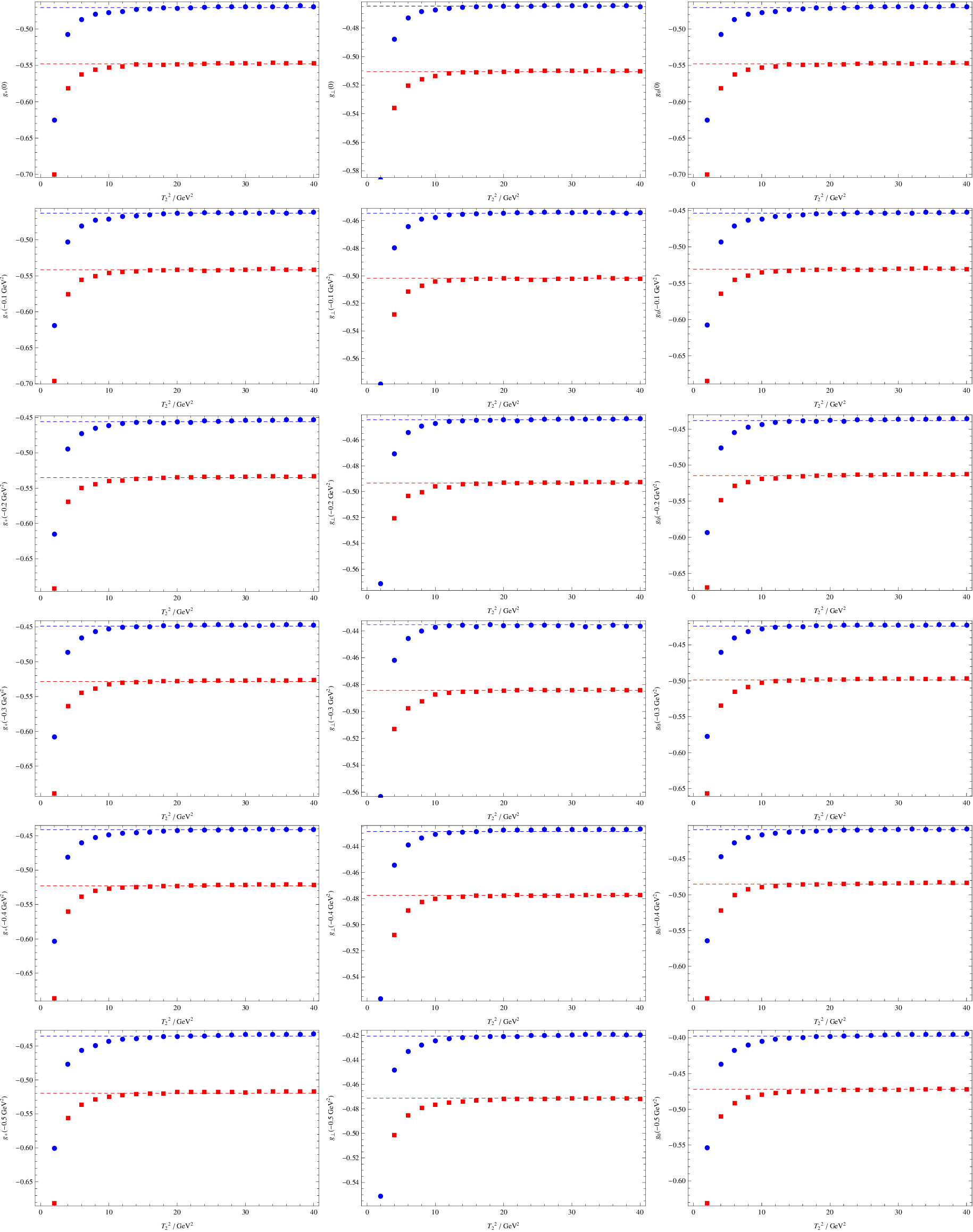}
\caption{Same as Fig. \ref{fig:FF_V}, but for $g_{+,\perp,0}$.}
\label{fig:FF_A} 
\end{figure}

In this subsection, the sum rule in Eq. (\ref{eq:Fi_plus_plus}) will
be investigated. As the most important parameters in QCDSR, the
continuum threshold parameters $s_{1,2}^{0}$ of initial and final
states, are taken from corresponding two-point
correlation functions, see Table \ref{Tab:pole_residues}. 

Reasonable Borel parameters should satisfy $T_{1}^{2}\sim{\cal O}(M_{1}^{2})$
and $T_{2}^{2}\sim{\cal O}(M_{2}^{2})$ with $M_{1,2}$ the masses
of initial and final baryons \cite{Shifman:1978bx,Ioffe:1981kw}.
In the following, we consider a line segment $T_{1}^{2}=3.5\ T_{2}^{2}$
with $T_{2}^{2}\in[2,40]\ {\rm GeV}^{2}$ on the $T_{1}^{2}-T_{2}^{2}$
plane. Relatively stable Borel windows can be found,
as can be seen in Figs. \ref{fig:FF_V} and \ref{fig:FF_A}. 

\begin{figure}[!]
\includegraphics[width=1.0\columnwidth]{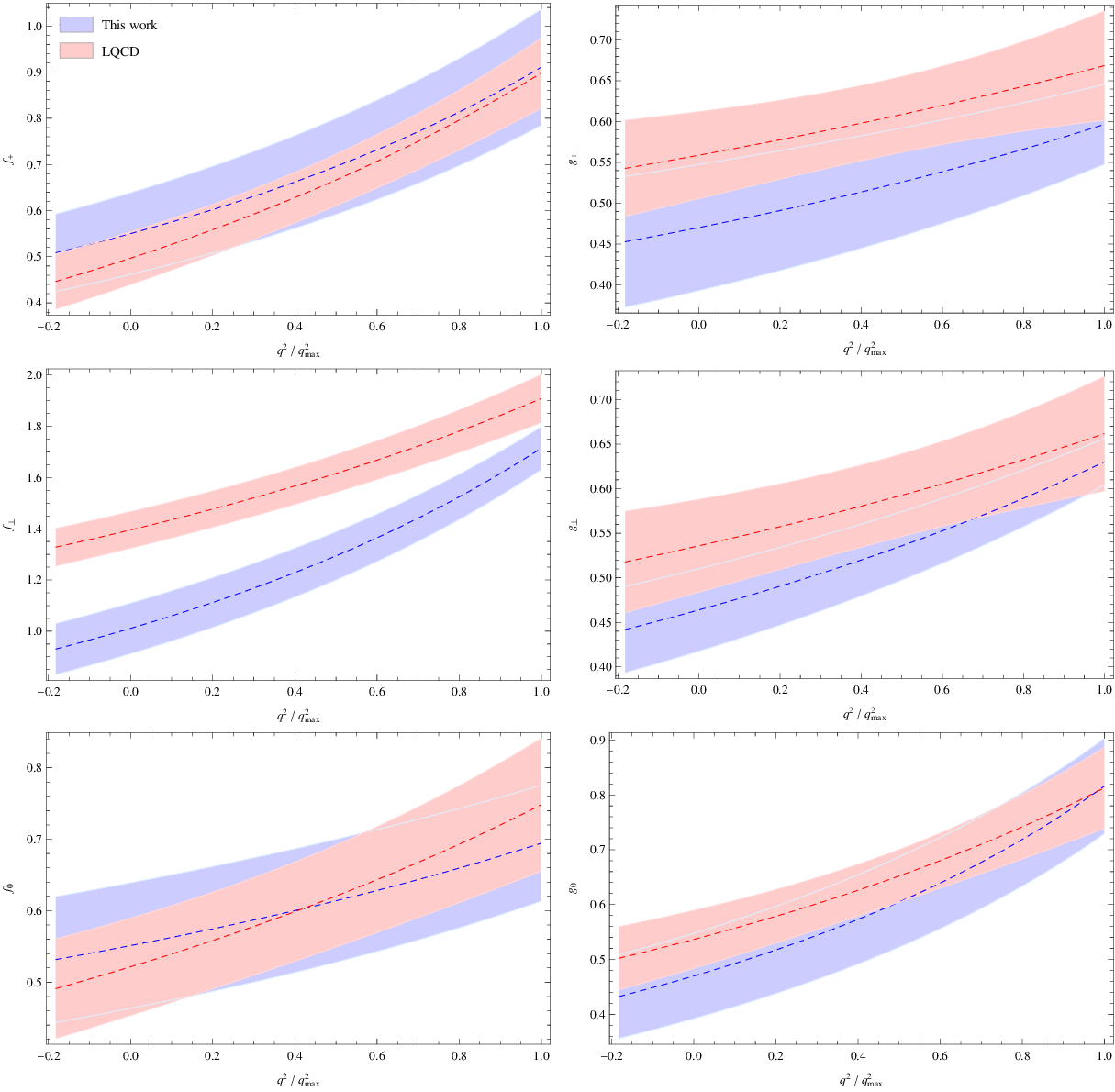}
\caption{Our helicity form factors are compared with those of Lattice QCD in Ref. \cite{Zhang:2021oja}.
All our form factors have been multiplied by a minus sign. }
\label{fig:FF_comparison} 
\end{figure}

To access the $q^{2}$ dependence of the $\Xi_{c}\to\Xi$ form factors,
we calculate the form factors for $q^{2}\in[-0.5,0]\ {\rm GeV}^{2}$, and
then fit the obtained values of $(q^{2},f(q^{2}))$ to the following
simplified $z$-expansion: 
\begin{equation}
f(q^{2})=\frac{a+b\ z(q^{2})}{1-q^{2}/(m_{{\rm pole}}^{f})^{2}},
\end{equation}
where
\begin{equation}
z(q^{2})=\frac{\sqrt{t_{+}-q^{2}}-\sqrt{t_{+}-t_{0}}}{\sqrt{t_{+}-q^{2}}+\sqrt{t_{+}-t_{0}}}
\end{equation}
with $t_{+}=(m_{D}+m_{K})^{2}$ and $t_{0}=q_{{\rm max}}^{2}=(m_{\Xi_{c}}-m_{\Xi})^{2}$.
The pole masses $m_{{\rm pole}}^{f}$ are respectively taken as $m_{{\rm pole}}^{f_{+},f_{\perp}}=2.112\ {\rm GeV}$,
$m_{{\rm pole}}^{f_{0}}=2.318\ {\rm GeV}$, $m_{{\rm pole}}^{g_{+},g_{\perp}}=2.460\ {\rm GeV}$,
and $m_{{\rm pole}}^{g_{0}}=1.968\ {\rm GeV}$ \cite{Zhang:2021oja}.
The fitted results of $(a,b)$ are given in Table \ref{Tab:a_b}.
In Fig. \ref{fig:FF_comparison}, our helicity form factors are compared with those
of Lattice QCD \cite{Zhang:2021oja}. One can see that, most of our
form factors are consistent with those
of Lattice QCD within error, except for $f_{\perp}$. 

\begin{table}
\caption{The fitted results of $(a,b)$ for the $\Xi_{c}\to\Xi$ form factors.}
\label{Tab:a_b} %
\begin{tabular}{c|c||c|c}
\hline 
$\overline{{\rm MS}}$ mass & $(a,b)$ & Pole mass@NLO & $(a,b)$\tabularnewline
\hline 
$f_{+}$ & $(-0.642,1.358)$ & $f_{+}$ & $(-0.730,1.362)$\tabularnewline
$f_{\perp}$ & $(-1.208,2.919)$ & $f_{\perp}$ & $(-1.267,2.329)$\tabularnewline
$f_{0}$ & $(-0.524,-0.402)$ & $f_{0}$ & $(-0.585,-0.796)$\tabularnewline
$g_{+}$ & $(-0.467,-0.050)$ & $g_{+}$ & $(-0.505,-0.628)$\tabularnewline
$g_{\perp}$ & $(-0.493,0.433)$ & $g_{\perp}$ & $(-0.514,0.050)$\tabularnewline
$g_{0}$ & $(-0.538,1.017)$ & $g_{0}$ & $(-0.596,0.714)$\tabularnewline
\hline 
\end{tabular}
\end{table}

\subsection{Phenomenological applications}

Our form factors are then used to predict the semileptonic decay widths.
The polarized decay widths for ${\cal B}_{1}\to{\cal B}_{2}l\nu$
are given by 
\begin{align}
\frac{d\Gamma_{L}}{dq^{2}} & =\frac{G_{F}^{2}|V_{{\rm CKM}}|^{2}q^{2}\ p\ (1-\hat{m}_{l}^{2})^{2}}{384\pi^{3}M_{1}^{2}}\left((2+\hat{m}_{l}^{2})(|H_{-\frac{1}{2},0}|^{2}+|H_{\frac{1}{2},0}|^{2})+3\hat{m}_{l}^{2}(|H_{-\frac{1}{2},t}|^{2}+|H_{\frac{1}{2},t}|^{2})\right),\label{eq:decay_width_L}\\
\frac{d\Gamma_{T}}{dq^{2}} & =\frac{G_{F}^{2}|V_{{\rm CKM}}|^{2}q^{2}\ p\ (1-\hat{m}_{l}^{2})^{2}(2+\hat{m}_{l}^{2})}{384\pi^{3}M_{1}^{2}}(|H_{\frac{1}{2},1}|^{2}+|H_{-\frac{1}{2},-1}|^{2}).\label{eq:decay_width_T}
\end{align}
where $\hat{m}_{l}\equiv m_{l}/\sqrt{q^{2}}$, and $p=\sqrt{Q_{+}Q_{-}}/(2M_{1})$
with $Q_{\pm}=(M_{1}\pm M_{2})^{2}-q^{2}$, $M_{1}=m_{\Xi_{c}}$,
$M_{2}=m_{\Xi}$. The helicity amplitudes $H_{\lambda_{2},\lambda_{W}}\equiv H_{\lambda_{2},\lambda_{W}}^{V}-H_{\lambda_{2},\lambda_{W}}^{A}$,
where $H_{\lambda_{2},\lambda_{W}}^{V,A}$ can be written in terms of the helicity form factors
\begin{align}
 & H_{\frac{1}{2},0}^{V}=-i\frac{\sqrt{Q_{-}}}{\sqrt{q^{2}}}(M_{1}+M_{2})f_{+},\quad H_{\frac{1}{2},1}^{V}=-i\sqrt{2Q_{-}}f_{\perp},\quad H_{\frac{1}{2},t}^{V}=-i\frac{\sqrt{Q_{+}}}{\sqrt{q^{2}}}(M_{1}-M_{2})f_{0},\nonumber \\
 & H_{\frac{1}{2},0}^{A}=-i\frac{\sqrt{Q_{+}}}{\sqrt{q^{2}}}(M_{1}-M_{2})g_{+},\quad H_{\frac{1}{2},1}^{A}=-i\sqrt{2Q_{+}}g_{\perp},\quad H_{\frac{1}{2},t}^{A}=-i\frac{\sqrt{Q_{-}}}{\sqrt{q^{2}}}(M_{1}+M_{2})g_{0},
\end{align}
and 
\begin{equation}
H_{-\lambda_{2},-\lambda_{W}}^{V}=H_{\lambda_{2},\lambda_{W}}^{V},\quad H_{-\lambda_{2},-\lambda_{W}}^{A}=-H_{\lambda_{2},\lambda_{W}}^{A}.
\end{equation}

Finally, we arrive at: 
\begin{align}
{\cal B}(\Xi_{c}^{0}\to\Xi^{-}e^{+}\nu_{e}) & =(1.83\pm0.45)\%,\nonumber \\
{\cal B}(\Xi_{c}^{0}\to\Xi^{-}\mu^{+}\nu_{\mu}) & =(1.77\pm0.43)\%,\nonumber \\
{\cal B}(\Xi_{c}^{+}\to\Xi^{0}e^{+}\nu_{e}) & =(5.58\pm1.36)\%,\nonumber \\
{\cal B}(\Xi_{c}^{+}\to\Xi^{0}\mu^{+}\nu_{\mu}) & =(5.38\pm1.31)\%,\label{eq:Brs}
\end{align}
where $\tau(\Xi_{c}^{+})=(453\pm5)$ fs and $\tau(\Xi_{c}^{0})=(151.9\pm2.4)$
fs have been used \cite{ParticleDataGroup:2022pth}. The central values
are obtained using the $\overline{{\rm MS}}$ scheme for the
quark masses, and the uncertainties are obtained by further considering
the pole mass scheme at NLO. One can see from Eq. (\ref{eq:Brs}) that,
the NLO corrections for the branching ratios may be around $25\%$.
In addition, Eq. (\ref{eq:Brs}) leads to 
\begin{equation}
{\cal B}(\Xi_{c}^{0}\to\Xi^{-}e^{+}\nu_{e})/{\cal B}(\Xi_{c}^{0}\to\Xi^{-}\mu^{+}\nu_{\mu})=1.037\pm0.002,
\end{equation}
which is in perfect agreement with the experimental value $1.03\pm0.05\pm0.07$ obtained by Belle \cite{Li:2021uhk}. 

Considering the lifetime of $\Xi_{c}^{0}$ changing from around 112
fs in PDG2018 \cite{Tanabashi:2018oca} to around 152 fs in PDG2022
\cite{ParticleDataGroup:2022pth}, in Table \ref{Tab:comparison_semi_lep},
only the decay width is compared with those from other theoretical
predictions and experimental measurements. It can be seen that, most
theoretical predictions are larger than the most recent experimental
data from Belle, and our result is consistent with those of ALICE
and Belle, and that of Lattice QCD.

\begin{table}
\caption{Our decay width of $\Xi_{c}\to\Xi e^{+}\nu_{e}$ (in units of
$10^{-13}$ GeV) is compared with experimental data,
and other theoretical predictions including Lattice QCD (LQCD),
light cone sum rules (LCSR), light-front quark model (LFQM),
relativistic quark model (RQM), and SU(3) flavor symmetry (SU(3)). }
\label{Tab:comparison_semi_lep} 
\begin{tabular}{c|c|c|c|c|c}
\hline 
This work  & LCSR \cite{Duan:2022yia}  & LFQM \cite{Ke:2021pxk} & LCSR \cite{Aliev:2021wat} & LCSR \cite{Azizi:2011mw}  & SU(3) \cite{Geng:2019bfz} \tabularnewline
\hline 
$0.79\pm0.19$  & $1.21\pm0.07$  & $0.74\pm0.15$  & $0.80\pm0.24$  & $4.26\pm1.49$  & $1.6\pm0.1$ \tabularnewline
\hline 
\end{tabular}
\begin{tabular}{c|c|c|c|c}
\hline 
RQM \cite{Faustov:2019ddj} & LFQM \cite{Zhao:2018zcb} & LQCD \cite{Zhang:2021oja} & ALICE \cite{ALICE} & Belle \cite{Li:2021uhk}\tabularnewline
\hline 
$1.40$ & $0.80$ & $1.02\pm0.19$ & $1.04\pm0.36$ & $0.563\pm0.168$\tabularnewline
\hline 
\end{tabular}
\end{table}

\section{Conclusions}

In this work, the $\Xi_{c}\to\Xi$ form factors are
investigated in QCD sum rules. To this end, the two-point correlation functions
of $\Xi_{c}$ and $\Xi$, and the three-point correlation functions of $\Xi_{c}\to\Xi$
are calculated. At the QCD level, contributions from up to dimension-6 four-quark
operators are considered, and the leading order results of the Wilson coefficients are obtained.
As the most important parameters in the
calculation of form factors, the continuum threshold parameters of $\Xi_{c}$ and $\Xi$
are determined using the derived sum rule for baryon mass.
For the form factors, relatively stable Borel windows can be found.
In this sense, our entire calculation has almost no adjustable parameters.

To reduce the scale dependence of our results, the leading logarithmic approximation is considered.
To roughly estimate the contribution from the next-to-leading order, we also take the quark masses as
the pole masses, and expand them to the next-to-leading order.
The corresponding results are then compared with those obtained in the $\overline{{\rm MS}}$ scheme.
Finally, our form factors are then used to predict the branching ratios of $\Xi_{c}\to\Xi\ell^{+}\nu_{\ell}$,
and we find that the next-to-leading order corrections for the branching ratios may be around $25\%$.
Our predictions of the branching ratios are consistent with those of ALICE and Belle, and that of Lattice QCD. 

In fact, a branching ratio itself is not enough for precise
comparison between theory and experiment. The form factors contain
more information! We suggest the experimentalists directly measure
the form factors of $\Xi_{c}\to\Xi\ell^{+}\nu_{\ell}$, and we believe
that our work will also help resolve the tension between the recent
Lattice QCD simulation and Belle's measurement. 

\section*{Acknowledgements}

The author is grateful to Profs. Pietro Colangelo, Wei Wang, Fu-Sheng Yu, and Drs. Yu-Ji
Shi, Yu-Shan Su, Qi-An Zhang for valuable discussions, and in particular, the author
would like to thank Prof. Wang Wei for his constant help and encouragement.
This work is supported in part by scientific research start-up fund
for Junma program of Inner Mongolia University, scientific research
start-up fund for talent introduction in Inner Mongolia Autonomous
Region, and National Natural Science Foundation of China under Grant
No. 12065020.

\end{document}